\documentstyle[11pt,appb,epsf]{article} 
\newcommand{\nll}{\nonumber \\}

\newcommand{\bq}{\begin{equation}}
\newcommand{\eq}{\end{equation}}
\newcommand{\ba}{\begin{eqnarray}}
\newcommand{\ea}{\end{eqnarray}}

\newcommand{\nobody}{\rule{0ex}{1ex}}

\newcommand{\nobodyfrac}{\frac{\nobody}{\nobody}}
\newcommand{\req}[1]{(\ref{#1})}
\begin{document}
\title{
\nobody\vspace{-2cm}\hfill\\ LMU-14/97\hfill\vspace{1cm}\\
Form factors and radiative Corrections in $Z'$ Physics\thanks
{Lecture given at the XXI school of Theoretical Physics of the Katowice
Univ., ``Recent progress in theory and phenomenology of fundamental
interactions'', Ustro\'n, September 97}
% you can use \\ to break lines
}
\author{
A. Leike
\address{
Ludwigs--Maximilians-Universit\"at, Sektion Physik, Theresienstr. 37,\\
D-80333 M\"unchen, Germany,\\
E-mail: leike@graviton.hep.physik.uni-muenchen.de
}
}
\maketitle
\begin{abstract}
This lecture contains a pedagogical approach to the description of
$Z'$ physics in the formalism of form factors.
Usually, only electroweak corrections are described by form factors, which
modify the Weinberg angle and the overall normalization.
It is demonstrated how this formalism can be extended
to include different Born contributions.
In the second part of the lecture, QCD and QED corrections are considered.
The development and consequences of the radiative tail for a $Z'$
search are discussed in detail.
\end{abstract}
\section{Introduction}
%--------------------------
The existence of an extra neutral gauge boson ($Z'$) is predicted in
many theories \cite{e6,cvetrev}, which go beyond the Standard Model (SM).
Although the mass of the $Z'$ is generally not predicted, there are
theories where its mass is naturally of the order of the electroweak
breaking scale \cite{9707451}.
If a $Z'$ is found at present or future colliders, it will provide us
with information on the large gauge group and on its symmetry breaking.
A $Z'$ search is therefore foreseen at every present and future
collider.

In this lecture, we focus on two topics.
The first topic is the description of $Z'$ physics in terms of
form factors of the $Z$ \cite{trunpub}.
This formalism has the advantage that it appears at the level of the
amplitude. 
It is therefore applicable equally well to all four fermion interactions.
These processes would give the main information on a $Z'$ at
$e^+e^-,\ e^\pm p,\ pp(p\bar p)$ colliders.
Form factors allow an easy extension of existing SM programs to $Z'$ physics. 

The second topic are QCD and QED corrections. 
They are known to be very process dependent.
The radiative tail arising due to initial state radiation is discussed
in detail for the process $e^+e^-\rightarrow f\bar f$.

Specific $Z'$ models \cite{e6,cvetrev} and present experimental bounds
\cite{9707451,sr97talk} on the $Z'$ mass and the $ZZ'$ mixing angle are not
discussed in this lecture. 
See \cite{leike97proc} for simple estimates of these bounds.

The form factors as known from electroweak corrections are introduced
in section 2.
They are applied to different Born contributions relevant in $Z'$ processes in
section 3.  
In section 4, radiative corrections are discussed.
Much room is devoted to the discussion of the radiative tail.
We conclude in section 5. 
\section{Form Factors in four Fermion Interactions}
%--------------------------
\subsection{Assumptions and Definitions}
%--------------------------
We will consider only the case of massless final fermions.
Observables, which are sensitive to transversal polarizations will not
be discussed.
Then, one can investigate the unpolarized case.
Different helicity states can be obtained by a simple substitution of
the couplings.
Consider the four fermion interaction $e^+e^-\rightarrow f\bar f\
(f\neq e)$ as an example.

The SM $Z$ boson couples to all known fermions $f$ with non--zero vector
and axial vector couplings $v_f,a_f$.
Denote the amplitude mediated by the SM $Z$ boson by ${\cal M}^Z$,
\bq
\label{fourfampl}
{\cal M}^Z= \frac{g_Z^2}{s-m_Z^2}
\bar{v}(e)\gamma_\beta \left[v_e-\gamma_5 a_e\nobodyfrac\right] u(e)\cdot
\bar{u}(f)\gamma^\beta \left[v_f -\gamma_5 a_f\nobodyfrac\right] v(f).
\eq
$m_Z$ stands for the complex mass of the $Z$ boson, 
$m_Z^2=M_Z^2-i\Gamma_ZM_Z$,
$g_Z^2$ is the coupling strength of the fermion current to the
$Z$ boson, and $s$ is the c.m. energy.

In general, it is not sufficient to describe four fermion
interactions by the $Z$ amplitude alone.
Suppose that an extra amplitude ${\cal M}$ must be added to ${\cal
M}^Z$ to reach a given precision,
\ba
\label{fourfampl2}
{\cal M}^\Sigma&=&{\cal M}^Z+{\cal M}\nll
&\equiv& \frac{g_Z^2}{s-m_Z^2}                            
\bar{v}(e)\gamma_\beta u(e)\cdot\bar{u}(f)\gamma^\beta v(f)\cdot v_e(1)v_f(1)
(1+\epsilon_{vv}) \\
&&\hspace{1.0cm}
-\bar{v}(e)\gamma_\beta u(e)\cdot\bar{u}(f)\gamma^\beta\gamma_5v(f)
\cdot v_e(1)a_f(1)(1+\epsilon_{va}) \nll
&&\hspace{1.0cm}
-\bar{v}(e)\gamma_\beta\gamma_5u(e)\cdot\bar{u}(f)\gamma^\beta v(f)
\cdot a_e(1)v_f(1)(1+\epsilon_{av}) \nll
&&\hspace{1.0cm}
+\bar{v}(e)\gamma_\beta\gamma_5u(e)\cdot\bar{u}(f)\gamma^\beta\gamma_5v(f)
\cdot a_e(1)a_f(1)(1+\epsilon_{aa}).\nonumber
\ea
The extra amplitude can be large but it must have the same structure
as ${\cal M}^Z$.
Then, the contribution of ${\cal M}$ can be absorbed into ${\cal
M}^Z$ by a redefinition of the $Z$ couplings to fermions,
\bq
\tilde x_e\tilde y_f=x_ey_f (1+\epsilon_{xy}),\ \ \ x,y=a,v.
\eq
The coefficients $\epsilon_{xy}$ in equation \req{fourfampl2} contain
all information on the amplitude ${\cal M}$.
Several examples of $\epsilon_{xy}$ will be discussed in the next sections.

Following the tradition of electroweak corrections \cite{elweak},
the contributions $\epsilon_{xy}$ are parametrized by (complex) form factors
$\rho_{ef}$, $\kappa_e$, $\kappa_f$, $\kappa_{ef}$, which are introduced by
replacements of the couplings \cite{formfactors}, 
\ba
\label{formrepl}
v_ev_f&\rightarrow& a_e a_f
\left[\nobodyfrac 1-4|Q^e|s^2_W\kappa_e
-4|Q^f|s^2_W\kappa_f + 16|Q^eQ^f|s^4_W\kappa_{ef}\right],\nll
v_e&\rightarrow& a_e\left[\nobodyfrac 1-4|Q^e|s^2_W\kappa_e\right],\nll
v_f&\rightarrow& a_f\left[\nobodyfrac 1-4|Q^f|s^2_W\kappa_f\right],\nll
a_e,a_f&\rightarrow&\mbox{unchanged},\nll
g_Z^2=&\rightarrow& g_Z^2\rho_{ef}.
\ea
$s_W^2=\sin^2\theta_W$ and $\theta_W$ is the Weinberg angle.
Comparing with equation \req{fourfampl2}, the form factors can be expressed
through $\epsilon_{xy}$, 
\ba
\label{fourfampl3}
\rho_{ef}&=&1+\epsilon_{aa}\\
\kappa_f&=&\frac{1}{\rho_{ef}}
\left[1+\frac{\epsilon_{av}v_f-\epsilon_{aa}a_f}
{v_f-a_f}\right],\nll 
\kappa_e&=&\frac{1}{\rho_{ef}}
\left[1+\frac{\epsilon_{av}v_e-\epsilon_{aa}a_e}
{v_e-a_e}\right]\nll
\kappa_{ef}&=&\frac{1}{\rho_{ef}}
\left[1+\frac{\epsilon_{vv}v_ev_f+\epsilon_{aa}a_ea_f
             -\epsilon_{av}a_ev_f-\epsilon_{va}v_ea_f}
{[v_e-a_e][v_f-a_f]}\right].\nonumber
\ea
If the relation
\bq
\label{factorize}
(1+\epsilon_{va})(1+\epsilon_{av})=(1+\epsilon_{aa})(1+\epsilon_{vv})
\eq
is fulfilled, we have $\kappa_{ef}=\kappa_e\kappa_f$, i.e. the product
$v_ev_f$ needs no special replacement rule in \req{formrepl}.
\subsection{Sum Rule for Form Factors}
%--------------------------
Consider the case where two additional amplitudes are added to ${\cal M}^Z$,
\bq
\label{fourfampl4}
{\cal M}^\Sigma = {\cal M}^Z + {\cal M}^1 +{\cal M}^2.
\eq 
Suppose that the form factors
$\rho_{ef}^i,\kappa_e^i,\kappa_f^i,\kappa_{ef}^i,\ i=1,2$ are known for both
additional amplitudes ${\cal M}^1$ and ${\cal M}^2$.
Then, the combined form factors can be calculated taking into account
that $\epsilon_{xy}^\Sigma=\epsilon_{xy}^1+\epsilon_{xy}^2$,
\ba
\label{fourfampl5}
\rho_{ef}^\Sigma&=&\rho_{ef}^1+\rho_{ef}^2-1,\nll
\kappa_f^\Sigma&=&\frac{\kappa_f^1\rho_{ef}^1+\kappa_f^2\rho_{ef}^2-1}
{\rho_{ef}^1+\rho_{ef}^1-1},\nll
\kappa_{ef}^\Sigma&=&\frac{\kappa_{ef}^1\rho_{ef}^1+\kappa_{ef}^2\rho_{ef}^2-1}
{\rho_{ef}^1+\rho_{ef}^1-1}.
\ea
The sum rules \req{fourfampl5} are exact.
In many applications, the form factors are not very different from one.
Then, the approximate sum rules
\bq
\label{fourfampl6}
\rho_{ef}^\Sigma =\rho_{ef}^1\rho_{ef}^2,\ \ \ 
\kappa_f^\Sigma =\kappa_f^1\kappa_f^2,\ \ \ 
\kappa_{ef}^\Sigma =\kappa_{ef}^1\kappa_{ef}^2
\eq
can be used. In their derivation, contributions proportional to 
$\epsilon_{xy}\epsilon_{x'y'}$ are neglected.
\section{Born Contributions in Terms of Form Factors}
%--------------------------
\subsection{Photon Exchange}
%--------------------------
If we are not dealing with pure neutrino processes, the
complete Born amplitude in the SM consists of contributions of $Z$ and
photon exchange.
The photon exchange can be treated as the additional matrix
element in equation \req{fourfampl2}
\bq
\label{gammaampl}
{\cal M}={\cal M}^\gamma= \frac{e^2}{s}
\bar{v}(e)\gamma_\beta Q^eu(e)\cdot \bar{u}(f)\gamma^\beta Q^f v(f).
\eq
Comparing equations \req{gammaampl} and \req{fourfampl2}, we get
\bq
\epsilon_{av}^\gamma=\epsilon_{va}^\gamma=\epsilon_{aa}^\gamma=0\ \ \ 
\epsilon_{vv}^\gamma=\frac{e^2(s-m_Z^2)}{g_Z^2s}\frac{Q^eQ^f}{v_ev_f}.
\eq
The form factors follow from equation \req{fourfampl3},
\bq
\label{formgamma}
\rho_{ef}^\gamma=\kappa_e^\gamma=\kappa_f^\gamma=1,\ \ \ 
\kappa_{ef}^\gamma=1
+\frac{e^2(s-m_Z^2)}{g_Z^2s}\frac{Q^eQ^f}{[v_e-a_e][v_f-a_f]}.
\eq
\subsection{$Z'$ Exchange}
%--------------------------
The description of four fermion interactions in theories with
extra neutral gauge bosons demands a third matrix element at the tree level,
\bq
\label{zpampl}
{\cal M}= {\cal M}^{Z'}= \frac{g_{Z'}^2}{s-m_{Z'}^2}
\bar{v}(e)\gamma_\beta \left[v'_e-\gamma_5 a'_e\nobodyfrac\right] u(e)\cdot
\bar{u}(f)\gamma^\beta \left[v'_f -\gamma_5 a'_f\nobodyfrac\right] v(f).
\eq
$v'_f(a'_f)$ stands for the vector (axial vector) coupling of the $Z'$ to
the fermion $f$, $m_{Z'}$ is the complex mass of the $Z'$ and $g_{Z'}$
denotes the coupling strength of the fermion current to the $Z'$.
The functions $\epsilon_{xy}^{Z'}$ follow immediately,
\bq
\epsilon_{xy}^{Z'}=\chi_{Z/Z'}\frac{x'_ey'_f}{x_ey_f},\ \ \ x,y=a,v,
\ \ \  
\chi_{Z'/Z}=\frac{g_{Z'}^2(s-m_Z^2)}{g_Z^2(s-m_{Z'}^2)}.
\eq
Again, the resulting form factors can be calculated using equation
\req{fourfampl3}, 
\ba
\label{formzp}
\rho_{ef}^{Z'}&=& 
1+\chi_{Z'/Z}\frac{a'_ea'_f}{a_ea_f},\nll
\kappa_e^{Z'}&=&\frac{1}{\rho_{ef}^{Z'}}
\left[1+\chi_{Z'/Z}\frac{[v'_e-a'_e]a'_f}
                        {[v_e-a_e]a_f}\right],\nll
\kappa_f^{Z'}&=&\frac{1}{\rho_{ef}^{Z'}}
\left[1+\chi_{Z'/Z}\frac{[v'_f-a'_f]a'_e}
                        {[v_f-a_f]a_e}\right],\nll
\kappa_{ef}^{Z'}&=&\frac{1}{\rho_{ef}^{Z'}}
\left[1+\chi_{Z'/Z}\frac{[v'_e-a'_e][v'_f-a'_f]}
                        {[v_e-a_e][v_f-a_f]}\right].
\ea
See reference \cite{zfitter} for an earlier derivation of these formulae.
We emphasize that the form factors \req{formgamma} and \req{formzp}
include the photon and $Z'$ contributions without any approximation. 
Of course, they are $s$-dependent, in general not small and even
singular (resonating) for $s\rightarrow 0\ (s\approx M_{Z'}^2)$.
\subsection{$ZZ'$ Mixing}
%--------------------------
There are no quantum numbers, which forbid a mixing between the $Z$
and the $Z'$.

Indeed, the mass matrix of the $Z$ and $Z'$ receives in the general
case non-diagonal entries $\delta M^2$  related to the vacuum expectation
values of the Higgs fields, 
%----------------------------------------------------------------------
\bq
\label{zznomix}
{\cal L}_M = \frac{1}{2}(Z,\ Z') 
\left( \begin{array}{rl}  M_Z^2       & \delta M^2 \\
                          \delta M^2  & M_{Z'}^2 \end{array} \right)
\left( \begin{array}{c} Z\\ Z'\end{array} \right).
\eq
%----------------------------------------------------------------------
The mass eigenstates $Z_1$ and $Z_2$ are connected with the
symmetry eigenstates $Z$ and $Z'$ by a mixing matrix,
\bq
\label{zzmix}
\left( \begin{array}{c} Z_1 \\ Z_2 \end{array} \right)
=
\left( \begin{array}{rl}  c_M & s_M \\
                        - s_M & c_M \end{array} \right)
\left( \begin{array}{c} Z \\ Z' \end{array} \right),
\eq
which depends on the mixing angle $\theta_M$ with $c_M=\cos\theta_M,\
s_M=\sin\theta_M$. 

The masses $M_1$ and $M_2$ of the mass eigenstates $Z_1$ and $Z_2$ are
\bq
M_{1,2}^2 =\frac{1}{2}\left[M_Z^2+M_{Z'}^2\pm
\sqrt{(M_Z^2-M_{Z'}^2)^2+4(\delta M^2)^2}\right].
\eq
It follows  $M_1<M_Z<M_2$.
Per definition, $Z_1$ is the light mass eigenstate investigated at LEP\,1.

The mixing predicts the couplings $a_f(n),v_f(n),\ n=1,2$ of the mass
eigenstates, 
\ba
\label{vf1}
a_f(1) =  c_M a_f + \frac{g_{Z'}}{g_Z} s_M a'_f &&
v_f(1) =  c_M v_f + \frac{g_{Z'}}{g_Z} s_M v'_f,\nll
a_f(2) =  c_M a'_f - \frac{g_{Z}}{g_{Z'}} s_M a_f &&
v_f(2) =  c_M v'_f - \frac{g_{Z}}{g_{Z'}} s_M v_f.
\ea

The considered process is now given by amplitudes with photon,
$Z_1$ and $Z_2$ exchange.
Photon and $Z_2$ exchange can be treated as explained in the previous
sections.

Here, we deal with the $Z_1$ amplitude in presence of $ZZ'$ mixing,
\ba
\label{z1ampl}
{\cal M}&=& {\cal M}^M\\
&&\nobody\hspace{-1.3cm}=\frac{g_{Z}^2}{s-m_1^2}
\bar{v}(e)\gamma_\beta \left[v_e(1)-\gamma_5 a_e(1)\nobodyfrac\right] u(e)\cdot
\bar{u}(f)\gamma^\beta \left[v_f(1) -\gamma_5 a_f(1)\nobodyfrac\right] v(f).
\nonumber
\ea
$m_1$ is the complex mass of the $Z_1$.
The functions $\epsilon^M_{xy}$ are
\ba
\label{epsweak}
\epsilon_{xy}^M&=&\chi_{Z_1/Z}
\left(c_M+\frac{g_{Z'}}{g_Z}s_M\frac{x'_e}{x_e}\right)
\left(c_M+\frac{g_{Z'}}{g_Z}s_M\frac{y'_f}{y_f}\right)-1,\ \ \ x,y=a,v,\nll
\chi_{Z_1/Z}&=&\frac{s-m_Z^2}{s-m_1^2}.
\ea
Again, the resulting form factors can be calculated using equation
\req{fourfampl3},
\ba
\label{formmix}
\rho_{ef}^M&=&
\chi_{Z_1/Z}\left(c_M+\frac{g_{Z'}}{g_Z}s_M\frac{a'_e}{a_e}\right)
\left(c_M+\frac{g_{Z'}}{g_Z}s_M\frac{a'_f}{a_f}\right),\nll
\kappa_f^M&=&\frac{1+\frac{s_Mg_{Z'}}{c_Mg_Z}\frac{v'_f-a'_f}{v_f-a_f}}
                  {1+\frac{s_Mg_{Z'}}{c_Mg_Z}\frac{a'_f}{a_f}},\nll
\kappa_{ef}^M&=&\kappa_e^M\kappa_f^M.
\ea
In the expression for $\kappa_{ef}^M$, relation \req{factorize} is used.

In contrast to the previous sections, the $ZZ'$ mixing introduces a
shift of the $Z$ mass taken into account by the factor $\chi_{Z_1/Z}$.
As it should be, this normalization factor drops out
in the formulae for the $\kappa$'s.
The mass shift could be observed by charged current processes
verifying the relation $M_W^2/M_1^2>M_W^2/M_Z^2=1-s_W^2$.
At the Born level, the choice of $m_Z^2$ or $m_1^2$ in the $Z$--amplitude
\req{fourfampl} is a question of convention.

In reference \cite{zefit}, the amplitude ${\cal M}^Z$ in \req{fourfampl}
is redefined replacing $m_Z^2$ by $m_1^2$.
Then, the coefficients $\epsilon_{xy}^M$ have no factor $\chi_{Z_1/Z}$.
The results obtained in reference \cite{zefit} are reproduced by 
the form factors \req{formmix} in that case.
\section{Radiative Corrections}
%--------------------------
\subsection{Electroweak Corrections}
%--------------------------
Radiative corrections are necessary to meet the precise experimental
data by accurate theoretical predictions.

In the calculation of SM electroweak corrections \cite{elweak}, a huge
number of Feynman diagrams must be taken into account.
The result can always be expressed in terms of the four form factors
\cite{formfactors}  
\bq
\label{weakform}
\rho_{ef}^{ew},\ \kappa_e^{ew},\ \kappa_f^{ew},\ \kappa_{ef}^{ew}.
\eq
They are complex functions depending on all parameters
of the theory and on kinematical variables.
Without box contributions, the $\kappa$'s would factorize.
In many applications, the box contributions are negligible leading to
the approximate factorization
$\kappa_{ef}^{ew}\approx\kappa_e^{ew}\kappa_f^{ew}$.
See \cite{zfitter} for a review on the weak form factors and further
references. 

In a $Z'$ search on the $Z_1$ resonance , weak corrections and $ZZ'$ mixing
must be taken into account simultaneously.
Weak corrections apply to the symmetry eigenstate $Z$, which is
different from the mass eigenstate $Z_1$ in the case of a non--zero mixing.
The factor $\chi_{Z_1/Z}$ in equation \req{epsweak} can be expanded in
the difference $M_Z-M_1$,
\bq
\label{propapprox}
\chi_{Z_1/Z}\approx 1+\frac{2M_Z(M_1-M_Z)}{s-m_1^2}.
\eq
The experimental upper bound on this difference is better than
$150\,MeV$, compare \cite{hollik96}.
This is much smaller than the $Z$ width.
Approximating $\chi_{Z_1/Z}$ by 1, one makes a normalization error
of the order ``e.w. corrections $\cdot(M_1-M_Z)^2/\Gamma_Z^2$'', which is
completely negligible. 

Using the approximate sum rules \req{fourfampl6} to combine the form
factors of weak corrections \req{weakform} and $ZZ'$ mixing
\req{formmix}, one makes an error 
of the order ``e.w. corrections$\cdot\theta_M$'', which is
still much smaller than the e.w. corrections and therefore
negligible. 

The form factor $\rho_{ef}^{ew}$ arises through the replacement of the
coupling constant $g_Z^2$ by the muon decay constant,
\bq
g_Z^2=\frac{\pi\alpha}{s^2_Wc^2_W}\rightarrow
\sqrt{2}G_\mu M_Z^2\rho_{ef}^{ew}\equiv
\sqrt{2}G_\mu M_1^2\rho_{ef}^{ew}\rho_{mix}
\eq
with $\rho_{mix}=M_Z^2/M_1^2\approx 1+2(M_Z-M_1)/M_1$.
$\rho_{mix}$ takes into account the tree level mass shift.
Historically, this effect was numerically important \cite{zefit}.
It is marginal with the present experimental constraint on $M_Z-M_1$.

Theories predicting extra neutral gauge bosons always contain many
new particles.
These particles contribute to the ``weak corrections'' of the whole
theory.
All these contributions are neglected in present analyses.
As far as there are no hints of physics beyond the SM, the error due
to this ignorance is expected to be small.
\subsection{QCD Corrections}
%--------------------------
QCD corrections are present in fermion interactions involving quarks.
Even in the case of massless fermions at the tree level, the QCD (and
QED) corrections depend on fermion masses because they
regularize the infrared and collinear singularities.
QCD corrections are different in processes where two or four quarks
are involved.
We concentrate from now on to the process $e^+e^-\rightarrow f\bar f$.
$O(\alpha_s)$ corrections are present for $f=q$.

The corrections depend on the parameters of
the theory and on kinematic variables.
If one integrates over the whole phase space, the QCD corrections
reduce simply to a factor, which multiplies the total Born cross
section $\sigma_T^0$.
To order $O(\alpha_s)$, we have 
\bq
\label{qcdcorr}
\sigma_T^{QCD}=\left(1+\frac{\alpha_s}{\pi}\right)\sigma_T^0.
\eq

The $O(\alpha_s)$ QCD corrections can be obtained from the $O(\alpha)$
QED corrections by an implementation of color factors.
QCD and QED corrections are very process dependent. 
They can not be expressed in terms of form factors.
\subsection{QED Corrections}
%--------------------------
QED corrections are present in processes involving charged particles.

In the reaction $e^+e^-\rightarrow f\bar f,\ f\neq\nu$, QED corrections
can be separated into initial state radiation, final state radiation
and the interference between them.
All these corrections give different contributions to $CP$ even and
$CP$ odd parts of the cross section.
The corrections are rather involved functions of the kinematical
parameters \cite{666}.

For a $Z'$ search, the initial state correction is numerically most
important. 
For later reference, we give the initial state correction to the
total cross section for the case where all kinematical variables
except the photon energy are integrated out,
\bq
\label{convol}
\sigma_T^{ISR}(s) = [1 + S(\epsilon)]\sigma_T^0(s)
+\int_{\epsilon}^{\Delta}dv\;\sigma_T^0\left(\nobodyfrac s(1-v)\right)\;
H_T(v).
\eq
The flux function $H_T(v)$ describes the probability of the
emission of a photon with the energy $v\sqrt{s}$, where $\sqrt{s}$ the
c.m. energy.
To order $O(\alpha)$, it is \cite{convolt}
\ba
\label{hard}
H_{T}(v) &=& \bar{H}_{T}(v) + \frac{\beta_e}{v} = 
\beta_e\;\frac{1+(1-v)^2}{2v},\nll
\beta_e &=& \frac{2\alpha}{\pi}\;(Q^e)^2\;(L_e-1),\ \ \
L_e = \ln\frac{s}{m_e^2},\ \ \ Q^e = -1.
\ea
The quantity $[1 + S(\epsilon)]$ in equation \req{convol}
describes the Born term plus corrections due to soft  and virtual photons. 
To order $O(\alpha)$, we have
\ba
\label{soft}
S(\epsilon) = \bar{S} + \beta_e \ln\epsilon = 
\beta_e\;\left(\ln\epsilon + \frac{3}{4}\right) + \frac{\alpha}
{\pi}\;(Q^e)^2\;\left(\frac{\pi^2}{3}-\frac{1}{2}\right).
\ea
See reference \cite{zet} for analytical results of the integral
\req{convol} in presence of a $Z'$.
\subsection{The Radiative Tail}
%--------------------------

Starting from the convolution \req{convol}, the origin of the
radiative tail and its magnitude can be estimated.
We do this ignoring details of the radiator functions $S(\epsilon)$ and
$H_T^e(v)$. 
The $s'$ dependence of the Born cross section is
\ba
\label{decomp}
\sigma_T^0(s')&\approx&
\frac{1}{s'}\frac{s'}{s'-m_m^2}\frac{s'}{s'-m_n^{*2}}\nll
&=&\frac{s}{m_n^{*2}-m_m^2}\frac{1}{s}
\left[\frac{s'}{s'-m_n^{*2}}-\frac{s'}{s'-m_m^2}\right].
\ea
For $m=n$, the first factor  of the last expression becomes
\bq
\label{radtail0}
\frac{s}{m_n^{*2}-m_n^2}=-\frac{i}{2}\frac{M_n}{\Gamma_n}\frac{s}{M_n^2}.
\eq
This imaginary quantity will give contributions to the cross section
only, if it is met by another imaginary multiplier. 
It arises from the $v$--integration \req{convol} over the
remaining factors in equation \req{decomp}.
Keeping only the relevant term after partial fraction
decomposition, one gets
\ba
\label{radtail}
\sigma_T^{ISR}(s)-\sigma_T^0(s)&\approx&
\frac{i}{2}\frac{M_n}{\Gamma_n}\frac{s}{M_n^2}\frac{M_n^2}{s}
\int_0^\Delta d v\;\frac{1}{1-v-m_n^{*2}/s}\nll
&\approx&\frac{i}{2}\frac{M_n}{\Gamma_n}
\ln\frac{m_n^{*2}/s-1+\Delta}{m_n^{*2}/s-1}.
\ea
The real part of the argument of the logarithm  is negative for
 $s>M_n^2$ and $\Delta>1-M_n^2/s$.
The first condition demands that the c.m. energy must be larger then
the mass of the resonance. 
The second condition allows the radiation of
photons with an energy $\sqrt{s}\Delta=\sqrt{s}(1-M_n^2/s)$.
If both conditions are fulfilled, the whole expression in equation
\req{radtail} becomes real and the radiative tail develops.

For events of the radiative tail, the energy of the remaining $e^+e^-$ pair
equals $M_n$, i.e. it annihilates on top of the $Z_n$ resonance.
Fermion pair production {\it on} resonance is enhanced by a factor
$M_n^2/\Gamma_n^2$ relative to {\it off} resonance production.
The convolution \req{convol} tells us that we have to integrate over
all possible photon energies.
Resonance enhancement is only obtained for photons, which shift the
energy of the remaining $e^+e^-$ pair in the narrow interval $M_n\pm\Gamma_n$.
Therefore, the enhancement factor $M_n^2/\Gamma_n^2$ is effectively
multiplied by $\Gamma_n/M_n$ leading to a combined enhancement
$M_n/\Gamma_n$ of the radiative tail.
This is exactly what we have in equations \req{radtail0} and \req{radtail}.

The magnitude of the radiative tail can now be estimated restoring 
the missing factor in equation \req{radtail}.
We find from equation \req{hard} a factor  $\beta_e/2$ in limit of
large energies, and a multiplier 2 due to the second 
contribution in the difference in equation \req{decomp},
\bq
\label{radtail2}
\mbox{rad.\ tail\ }\approx \sigma_T^{ISR}(s)-\sigma_T^0(s)\approx
\sigma_T^0(s;n,n)\cdot\beta_e\frac{\pi}{2}\frac{M_n}{\Gamma_n}.
\eq
The factor $\beta_e$ contains $\alpha$ because the radiation of an
additional photon is a process of higher order, and $L_e=
\ln\frac{s}{m_e^2}$ giving an 
enhancement for photons radiated collinear to the beam.

\begin{figure}[tbh]
%-----------Fig.1-----------------------------
%\ \vspace{1cm}\hfill\\
\begin{center}
\begin{minipage}[t]{7.8cm} {
\begin{center}
\hspace{-1.7cm}
\mbox{
\epsfysize=7.0cm
\epsffile[0 0 500 500]{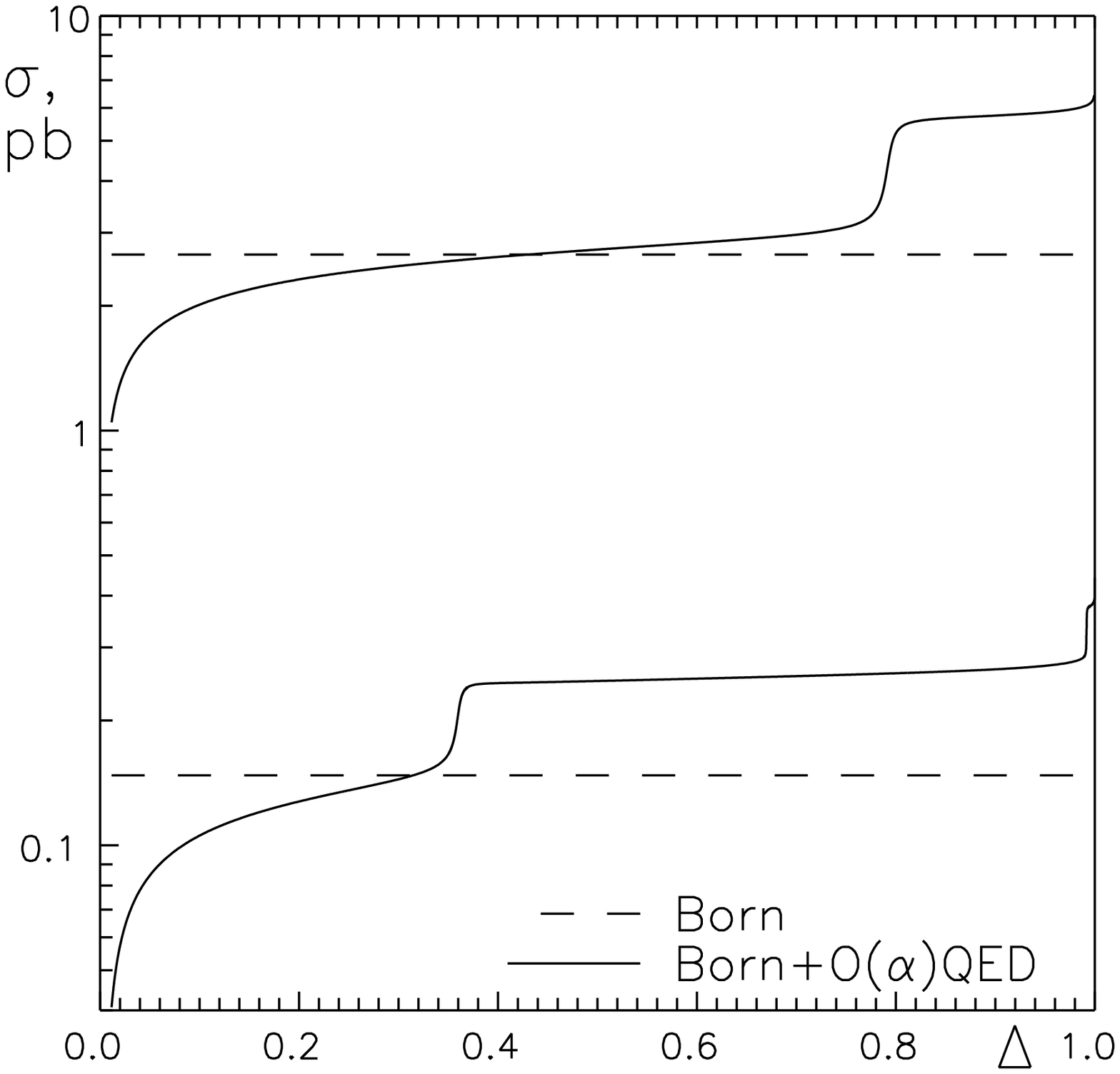}%
}
\end{center}
\vspace*{-0.5cm}
\noindent
{\small\it
{\bf Fig.1: }The total cross section $\sigma_T^\mu$ as function of the cut
on the photon energy $\Delta\sqrt{s}$ for $M_{Z'}=M_\eta=800\,GeV$. The upper
(lower) set of curves corresponds to $\sqrt{s}=200(1000)\,GeV$.
}
}\end{minipage}
\end{center}
\end{figure}

Putting $s_W^2=\frac{1}{4}$ and considering the $Z_1$ peak,
$M_n/\Gamma_n=M_1/\Gamma_1$, one gets  $\sigma_T^l(s;\gamma,Z_1)= 0$ and
$\sigma_T^l(s;Z_1,Z_1)/\sigma_T^l(s;\gamma,\gamma)= 1/9$
in the limit $s\gg M_Z^2$.
It follows 
\bq
\mbox{rad.\ tail\ }\approx 7\cdot\sigma_T^{l0}(s;Z_1,Z_1)\approx
7/10\cdot\sigma_T^{l0}, 
\eq
which is in reasonable agreement with the exact calculation and with
figure~1.  
For $b$ quark production where the $Z$ boson exchange
dominates over the photon exchange,
the effect of the radiative tail is much more pronounced;
$\sigma_T^{b0}(s;\gamma,Z_1)\approx 0$,
$\sigma_T^{b0}(s;Z_1,Z_1)/\sigma_T^{b0}(s;\gamma,\gamma)=4$ 
in the limit $s\gg M_Z^2$, and hence
\bq
\mbox{rad.\ tail\ }\approx 7\cdot\sigma_T^{b0}(s;Z_1,Z_1)\approx 28/5\cdot\sigma_T^{b0},
\eq
which is again in reasonable agreement with the exact calculation.

Only the cross section of the exchange of the vector boson $n$,
$\sigma_T^0(s;n,n)$, appears in equation \req{radtail2}.
All other contributions to $\sigma_T^0$ are not enhanced by the radiative tail.
For $M_1<\sqrt{s}<M_2$, the contribution $\sigma_T^0(s;Z_1,Z_1)$ is
enhanced, while the $Z'$ signal is not enhanced.
Therefore, the radiative tail must be removed for a $Z'$ search.
This can be done demanding $\Delta<1-M_1^2/s$.

The dependence of the cross section on $\Delta$
is shown in figure~1 for two different energies. 
The upper curves correspond to an energy above the $Z$ peak but below
the $Z'$ peak, the lower curves to an energy above the $Z$ and $Z'$ peaks.
One recognizes the step--like behaviour for photon energies where the
radiative tail(s) are ``switched on''. 
We see that the radiatively corrected cross section is numerically 
similar to the Born prediction only for a certain cut, 
which rejects all radiative tails.
This property is independent of the observable considered.
It is the reason why theoretical $Z'$ analyses with Born cross
sections agree very well with those involving SM corrections.
Of course, the whole SM corrections are needed for an analysis of real data.
\section{Conclusions}
%--------------------------
In this lecture, much attention was devoted to the implementation of
effects connected with $Z'$ physics by form factors.
This procedure works at the level of amplitudes.
Therefore, it can be applied to any four fermion interaction.
Examples are $e^+e^-\rightarrow f\bar f$, 
$e^+e^-\rightarrow e^+e^-$, $e^-e^-\rightarrow e^-e^-$,
$e^\pm q\rightarrow e^\pm q$,  $q\bar q\rightarrow f\bar f$.

QCD and QED corrections depend on the process
and on kinematical cuts.
Therefore, they have to be calculated for every reaction separately.
In the case of a $Z'$ search, the results known from the Standard
Model can be applied.

The radiative tail develops only in processes with resonances in the
$s$--channel. 
The mechanisms of its development and its magnitude are extensively discussed.
It is suppressed at hadron colliders mainly due to properties of the
structure functions.
At $e^+e^-$ colliders, it is most pronounced in fermion pair
production and therefore important at LEP\,2 and at a future linear collider.
For a $Z'$ search, it has to be removed by kinematical cuts.
\vspace{0.2cm}\\
\centerline{\bf Acknowledgement}
I would like to thank the organizing committee for the organization of
this stimulating school and the ``Stiftung f\"ur Deutsch--Polnische
Zusammenarbeit'' for financial support during the school.
%
%\vfill\newpage


\begin{thebibliography}{99}
\bibitem{e6} For a review see e.g.
             J.L. Hewett, T.G. Rizzo, Phys. Rep. {\bf 183} (1989) 193;\\
P. Langacker, M. Luo, A.K. Mann, Rev. Mod. Phys. {\bf 64} (1992) 87.
\bibitem{cvetrev}
A recent review can be found in
M. Cveti\v{c}, S. Godfrey, in ``Electroweak Symmetry Breaking and
Beyond the SM'', T. Barklow, S. Dawson, H. Haber, J. Siegrist (eds),
World Scientific 1995.
\bibitem{9707451} 
M. Cveti\v{c}, P. Langacker, hep-ph/9707451,
Univ. of Pensylvenia preprint UPR-0761-T;\\
M. Cveti\v{c}, P. Langacker, Phys. Rev. {\bf D54} (1996) 3570;\\
M. Cveti\v{c}, P. Langacker, hep-ph/9602424,
Univ. of Pensylvenia preprint UPR-0690-T;\\
J. R. Espinosa, Univ. Pensylvenia preprint UPR-0768-T, hep-ph/9707541.
\bibitem{trunpub} T. Riemann, 1991, unpublished.
\bibitem{sr97talk} S. Riemann, talk at Balholm, Norway, April 1997.
\bibitem{leike97proc} 
A. Leike, hep-ph/9708337, to appear in the Proceedings of the 
Meeting of the EC Network ``Electroweak Symmetry breaking'', 
Ouronapoulis, 1997;\\
A. Leike, hep-ph/9708436, to appear in DESY-123E/97.
\bibitem{elweak}
G. Passarino, M. Veltman, Nucl, Phys. {\bf B160} (1979) 151;\\
A.Akhundov, D. Bardin, T. Riemann, Nucl. Phys. {\bf B276} (1980) 1.
\bibitem{formfactors} 
A. Akhundov, D. Bardin, T. Riemann, Nucl. Phys. {\bf B276} (1986) 1;\\
D. Bardin et al., Z. Physik {\bf C44} (1989) 493;\\
G. Altarelli, R. Kleiss, C. Verzegnassi (eds.),
$Z$ physics at LEP 1, CERN 89-08 (1989) and references quoted therein.
\bibitem{zfitter} D. Bardin et al., CERN-TH.6443/92, hep-ph/9412201.
\bibitem{zefit} 
A. Leike, S. Riemann, T. Riemann, Phys. Lett. {\bf B291} (1992) 187;\\
A. Leike, S. Riemann, Nucl. Phys. {\bf B} (Proc. Suppl.) {\bf 29A} (1992) 270.
\bibitem{hollik96} W. Hollik, Acta Phys. Polon. {\bf B27} (1996) 3685.
\bibitem{666} D. Bardin, et al. , Nucl. Phys. {\bf B351} (1991) 1.
\bibitem{convolt} G. Bonneau, F. Martin, Nucl Phys. {\bf B27} (1971) 381.
\bibitem{zet} 
A. Leike, T. Riemann, M. Sachwitz, Phys. Lett. {\bf B241} (1990) 267.
\end{thebibliography}
\end{document}